\documentclass[twocolumn,showpacs,preprintnumbers,amsmath,amssymb]{revtex4}

\usepackage[dvips]{graphicx}

\usepackage{dcolumn}

\usepackage{bm}

\begin{document}

\title{Limitations in cooling electrons by normal metal - superconductor tunnel junctions}

\author{J. P. Pekola, T. T. Heikkil\"a, A. M. Savin, and J. T. Flyktman}
\affiliation{Low Temperature Laboratory, Helsinki University of
Technology, P.O. Box 3500, 02015 HUT, Finland}

\author{F. Giazotto}
\affiliation{NEST-INFM \& Scuola Normale Superiore, I-56126 Pisa, Italy}

\author{F. W. J. Hekking}
\affiliation{Laboratoire de Physique et Mod\'elisation des Milieux
Condens\'es, CNRS \& Universit\'e Joseph Fourier, BP 166, 38042
Grenoble-cedex 9, France}


\begin{abstract}
We demonstrate both theoretically and experimentally two limiting factors in cooling electrons using biased tunnel junctions to extract heat from a normal metal into a superconductor. 
Firstly, when the injection rate of 
electrons exceeds the internal relaxation rate in the metal to be cooled, the electrons do no more obey the Fermi-Dirac distribution, and the concept of temperature cannot be applied as 
such. Secondly, at low bath temperatures, states within the gap induce anomalous heating and yield a theoretical limit of the achievable minimum temperature.   
\end{abstract}

\maketitle

Refrigerators are generally characterised by their cooling power, coefficient of performance, and
operating temperature under various working conditions. To assign a temperature to a system, 
one needs to assume that the energy relaxation within the system is faster than any rate 
associated with the heat flux between the system in concern and its surroundings. If this 
condition fails, the energy distribution of the particles of which the system is 
formed is non-thermal, and applying the concept of temperature is strictly speaking inappropriate. Such a limit can be achieved in submicron-size coolers at low temperatures. The structure 
we study is a symmetric configuration of a NIS refrigerator \cite{nahum94}, formed by a series 
connection of two Superconductor (S) - Insulator (I) - Normal metal (N) tunnel junctions 
sharing the N island to be cooled in between them (SINIS) \cite{leivo96}. We demonstrate two striking phenomena occuring in these electron micro-coolers at low 
temperatures: evidence of non-thermal energy distribution of the cooled electrons and re-entrant behaviour with anomalous heating at low bias voltages. The cooler performance is typically 
limited by coupling of the electrons to the underlying lattice (electron-phonon, e-p coupling). This has, however, strong dependence on temperature $T$: relaxation rate $\tau _{\rm 
e-p}^{-1}$ slows down on lowering $T$ typically as $\tau _{\rm e-p}^{-1} \propto T^{3}$ \cite{roukes85}. Consequently, at low enough temperature, characteristically around 100 mK, the 
behaviour of the cooler can be described as if the lattice would not exist at all. The interplay of the rates for e-p, electron-electron (e-e) and the injection through the 
junctions determines the distribution in the normal metal. If e-p or e-e rates are fast, the system assumes Fermi-Dirac energy ($E$) 
distribution $f_0(E,T_e)$:
\begin{equation} \label{eq1}
\small{f_0(E,T_e) = \frac{1}{1+\exp(E/k_{\rm B}T_e)}}.
\end{equation}
In the limit of very strong e-p relaxation, $T_e$ equals the temperature 
of the lattice. We call this {\sl equilibrium}. On the other hand, if e-p relaxation is slow, and e-e relaxation is fast, $T_e$ is, in 
general, different from the lattice temperature ({\sl quasi-equilibrium}). Finally, the slow e-e relaxation as compared to the injection rates implies that the 
electrons assume a {\sl non-equilibrium} energy distribution $f(E)$, which cannot be written as Eq. (\ref{eq1}) \cite{pothier97,baselmans99}, and it is not possible to assign a true 
temperature to them.
\begin{figure}
\begin{center}
\includegraphics[width=0.45
\textwidth]{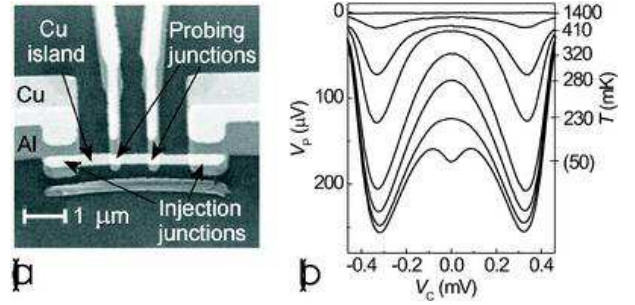}
\caption{Scanning electron micrograph of a typical cooler sample in (a), and cooling data in (b), where voltage $V_{\rm P}$ across the probe junctions in a constant current bias (28 pA) is 
shown against voltage $V_{\rm C}$ across the two injection junctions. Cryostat temperature, corresponding to the electron temperature on the N island at $V_{\rm C}=0$ is indicated on the 
right vertical axis. Yet below 100 mK this correspondence is uncertain, because of the lack of calibration and several competing effects to be discussed in the text.}\label{Figure 1}
\end{center}
\end{figure}

Our discussion is motivated by a puzzling experimental observation in many coolers at temperatures below or around 100 mK \cite{fisher99}. Figure 1 shows in (a) a typical 
SINIS cooler. The device has been fabricated by standard electron beam lithography. The central part forms the N island of copper (purity nominally 99.9999 \%). 
The injecting tunnel junctions, with normal-state resistances $R_{\rm T}$ (100 $\Omega$ - 2 k$\Omega$), contact the island symmetrically into the two superconducting 
aluminium reservoirs at the two ends of the N island. The overlapping extra copper shadows outside the SINIS structure provide better thermalisation of the Al reservoirs. Two additional NIS 
probe junctions, with normal-state resistances  $\gg 1$ k$\Omega$ in the centre are used to measure the 
temperature of the N island or to probe the distribution through the differential conductance of the nominally symmetric series connection. $R_{\rm T} \simeq$ 200 $\Omega$ and thickness of 
the copper film $\simeq$ 30 nm are the essential parameters of the sample (S1) whose data are presented in this paper.
The data in Fig. 1 (b) were taken by applying a constant current $I_0$ through the probe, which, due to the thermal rounding of the current-voltage characteristics, provides, by detecting 
voltage $V_{\rm P}$ across it, a measure of the electron temperature on the N island \cite{rowell76}. 
We calibrated this dependence by varying the bath temperature of the cryostat. Typically one applies $I_0$ such that the voltage remains within the gap region of the superconductor, $V_{\rm 
P} < 2\Delta/e$, whereby excess heating or cooling by the probes is 
avoided. Here $\Delta$ is the energy gap of the superconductor and $e$ is the electron charge ($\Delta /e \simeq 0.2$ mV). Figure 1 (b) shows $V_{\rm P}$ against injection voltage $V_{\rm 
C}$; the several curves represent different bath 
temperatures. The approximate temperature calibration is given on the 
right vertical axis. At all but the lowest bath temperature of about 50 mK, the curves show the expected refrigeration behaviour symmetrically around $V_{\rm C}=0$, with optimum cooling at 
about $V_{\rm C} \simeq \pm 2\Delta/e$ \cite{leivo96}. The lowest-temperature curve demonstrates, however, a feature which appears as heating at low values of $V_{\rm C}$ and re-entrant 
cooling again 
close to $2\Delta/e$. This behaviour is common with many similar samples, and it appears only below 200 mK. Moreover, the measured conductance curves of the probe junctions (Fig. 4) indicate 
that at low temperatures the actual temperature we assign in a $V_{\rm P}$ measurement depends on the choice of $I_0$.

To understand the observed behaviour, we consider the properties of the kinetic equation describing the dominant processes in our system. We assume that the two reservoirs are identical and 
the quasiparticles 
have a thermal distribution of Eq. (\ref{eq1}), with $T_e \equiv T_{\rm S}$ on them. In steady state $f(E)$ on the N island is then determined by \cite{andreevnote}
\begin{eqnarray} \label{eq2}
&& \frac{\delta}{e^2R_{\rm T}}{\big[}n(E_{\rm R})[f_0(E_{\rm R},T_{\rm S})-f(E)]\nonumber \\ && +n(E_{\rm L})[f_0(E_{\rm L},T_{\rm S})-f(E)] {\big]}=I_{\rm coll}[f;E].
\end{eqnarray}
Here $I_{\rm coll}[f;E]$ is the collision integral discussed below, $\delta$ is the level spacing on the island, $E_{\rm L,R}=E \pm eV_{\rm C}/2$ are energies on the left (L) and right (R) 
reservoirs, and
\begin{equation} \label{eq3}
n(E) = |{\rm Re}(\frac{E+i\Gamma}{\sqrt{(E+i\Gamma)^2-\Delta^2}})|
\end{equation}
is the broadened BCS density of states (DOS) of the superconductor. $\Gamma$ smears the DOS singularity at $E=\pm\Delta$ and allows for states 
within the gap, e.g., due to inelastic electron scattering in the superconductor \cite{dynes84} or by inverse proximity effect due to nearby normal metals. A more phenomenological choice is 
to add a non-zero constant 
$\Gamma/\Delta$ to the ideal singular DOS ($\Gamma \equiv 0$). This choice would not essentially affect our conclusions.  

It is illustrative to investigate first the case where relaxation (both e-e and e-p) is very weak, i.e., when $I_{\rm coll}[f;E]\equiv 0$ in Eq. (\ref{eq2}). Then we obtain an explicit 
expression for $f(E)$ as
\begin{equation} \label{eq4}
f(E)=\frac{n(E_{\rm R})f_0(E_{\rm R},T_{\rm S})+n(E_{\rm L})f_0(E_{\rm L},T_{\rm S})}{n(E_{\rm R})+n(E_{\rm L})}.
\end{equation}
The solution of Eq. (\ref{eq4}) is plotted in Fig. 2 (a) for five different values of $v_{\rm C} \equiv eV_{\rm C}/\Delta$, assuming $\eta \equiv \Gamma / \Delta= 1\cdot 10^{-4}$ 
\cite{eta} and $T_{\rm S}=0.1T_{\rm C}$. The BCS relation $\Delta \simeq 1.764 k_{\rm B} T_{\rm C}$ has been assumed for the critical temperature $T_{\rm C}$. This solution exhibits 
some nontrivial features. 
At low values of the bias voltage up to $v_{\rm C} \sim 1$, the distribution first broadens whereafter it starts to get narrower again, and at $v_{\rm C}=2.0$, it becomes very narrow, 
effectively demonstrating strong cooling. In each case, except at $v_{\rm C}=0$ (equilibrium), the distribution is not thermal. At $v_{\rm C} > 2.0$ the 
distribution would become even more unusual \cite{vanhuffelen93}, but we do not consider this regime in detail for reasons to be explained below. Yet, at those voltages the distribution 
effectively broadens again, suggesting already the non-monotonic behaviour, heating - cooling - heating upon increasing $v_{\rm C}$. 
\begin{figure}
\begin{center}
\includegraphics[width=0.5
\textwidth]{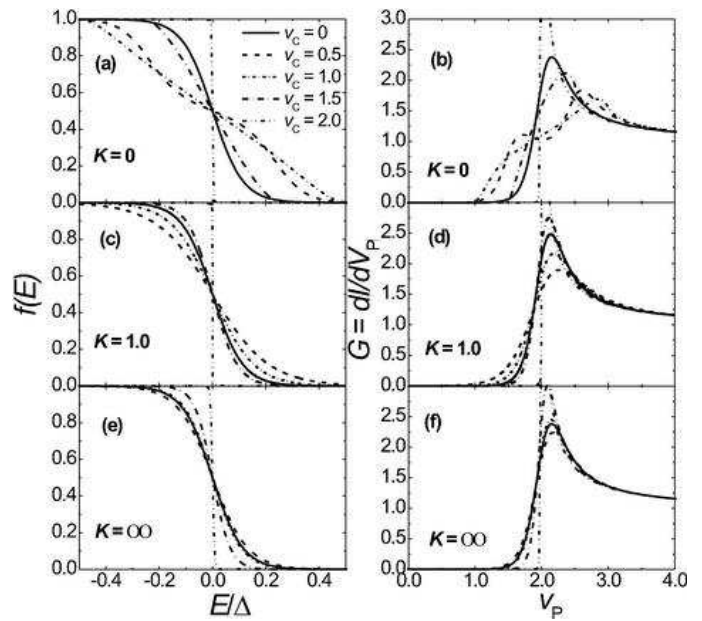}
\caption{Energy distribution $f(E)$ against $E/\Delta$ in (a), (c) and (e), and differential conductance $G$ against probe voltage $v_{\rm P}$ in (b), (d) and (f) at three different values 
of e-e collision parameter $K$, which present very slow, intermediate and very fast e-e relaxation from top to bottom, respectively. Parameter values $\eta = 1\cdot 10^{-4}$ and $T_{\rm 
S}/T_{\rm C}=0.1$ have been assumed.}\label{Figure 2}
\end{center}
\end{figure}
The distribution is tested by measuring the differential conductance $dI/dV_{\rm P}$ of the probe junctions (Fig. 1 (a)). The relation between $dI/dV_{\rm P}$, $V_{\rm P}$, and $f(E)$ is 
given by
\begin{equation} \label{eq5}
\frac{dI}{dV_{\rm P}} = - \int _{-\infty}^{\infty}n(E-eV_{\rm P}/2)\frac{df(E)}{dE}dE. 
\end{equation} 
Here $dI/dV_{\rm P}$ has been scaled by the normal-state conductance of the series connection of the probe junctions. 
$G \equiv \frac {dI}{dV_{\rm P}}$ has been plotted in Fig. 2 (b) for the collisionless distributions of Fig. 2 (a) against $v_{\rm P} \equiv eV_{\rm P}/\Delta$. These curves exhibit some 
similarities to those of biased diffusive metal wires \cite{pothier97}, especially at low values of $v_{\rm C}$, where quasi-constant DOS at low energies ($n(E)\simeq \Gamma/\Delta$) mimics 
resistive normal-metal wire. Another important limit is quasi-equilibrium with electrons perfectly decoupled from the phonon bath. The 
latter condition is well justified at low $T_{\rm S}$ by a standard estimate of the minimum temperature $T_{\rm min}$ determined by the balance between cooling 
power and e-p coupling only \cite{phononestimate}.
In quasi-equilibrium fast e-e relaxation forces the distribution into a thermal one (Eq. (\ref{eq1})). Any temperature satisfies this and we need to determine $T$ by 
setting the net heat flux from the island, $P(T,T_{\rm S})$, equal to zero. In our case we can write \cite{leivo96}
\begin{equation} \label{eq10}
\small{P(T,T_{\rm S})=\frac{2}{e^2R_{\rm T}}\int _{-\infty}^{\infty}n(E)[f_0(E_{\rm R},T)-f_0(E,T_{\rm S})]E_{\rm R} dE =0},
\end{equation}
where factor 2 on the right hand side takes into account the flux through the two identical junctions L and R. The solutions of Eq. (\ref{eq10}) allow us to plot the distributions and 
differential conductance at different values of $v_{\rm C}$ as in Fig. 2 (e) and (f). The re-entrant 
heating-cooling-heating behaviour survives but less pronounced than in the collisionless case. It transforms into more conventional cooling-heating 
behaviour above $T_{\rm S}=T^*$, given by
\begin{equation} \label{eq10b}
(\Delta/k_{\rm B}T^*)^{3/2}\exp(-\Delta/k_{\rm B}T^*) \simeq \eta/\sqrt{2\pi}.
\end{equation}
Thus $T^*$ depends approximately logarithmically on $\eta$ and assumes a value $T^*/T_{\rm C} \simeq 0.125$ when $\eta = 1 \cdot 10^{-4}$. Equation (\ref{eq10b}) is obtained by equating at 
low bias the quadratic in $v_{\rm C}$ heating and cooling terms arising from the states inside and outside the gap, respectively. For an intermediate strength of e-e interaction, the 
distribution
functions were obtained by Eq. (\ref{eq2}) with the e-e collision integral given
in Ref. \cite{pothier97}. In the case of a diffusive normal-metal island whose
transverse dimensions are smaller than the coherence length $\xi _0 =
\sqrt{\hbar D/\Delta}$, one finds $I_{\rm coll}=\kappa \sqrt{\Delta} \tilde{I}_{\rm coll}$ where
\begin{equation} \label{coll}
{\small
\begin{split}
\tilde{I}_{\rm coll}= &\int \frac{d\omega d\epsilon}{\omega^{3/2}}[f(E)f(\epsilon\Delta)(1-f(E-\omega\Delta))(1-f((\epsilon+\omega)\Delta))\\&
-f(E-\omega\Delta)f((\epsilon+\omega)\Delta)(1-f(E))(1-f(\epsilon\Delta))]
\end{split}}
\end{equation}
is the dimensionless collision integral, $\kappa=\frac{\sqrt{2} L \delta}{\pi \sqrt{D} \hbar^{3/2}}$, and $D$ and $L$ are the diffusion coefficient and the length of the island. Dividing Eq. 
(\ref{eq2}) by   
$\delta/(e^2 R_{\rm T})$, we obtain a dimensionless equation where 
the strength of e-e scattering is governed by
$K \equiv 2\sqrt{2}\frac{R_{\rm T}}{R_{\rm K}} \sqrt{\frac{\Delta}{E_{\rm T}}}$ \cite{collision}. Here $R_{\rm K} \equiv h/e^2$ and $E_{\rm T} \equiv \hbar D/L^2$ are the resistance quantum 
and the Thouless energy of the island, respectively.

Although temperature is not a valid concept in non-equilibrium, we can define an effective temperature $T_{\rm eff}$. A natural choice is to require that $T_{\rm eff}$ 
satisfies the standard relation between the temperature and the thermal energy density of electrons in Sommerfeld expansion, which yields
\begin{equation} \label{eq11}
k_{\rm B}T_{\rm eff}=\frac{\sqrt{6}}{\pi}\sqrt{\int _{-\infty}^{\infty}[f(E)-1+\theta(E)]E dE}.
\end{equation}
Here $\theta(E)$ is the Heaviside step function. This $T_{\rm eff}$ coincides with the true temperature in (quasi-)equilibrium, and it is not affected by the strength 
of e-e relaxation as such. 
Yet in a biased SINIS, $T_{\rm eff}$ depends on the strength of e-e relaxation, because of the heat exchange with reservoirs with non-constant DOS. 
\begin{figure}
\begin{center}
\includegraphics[width=0.35
\textwidth]{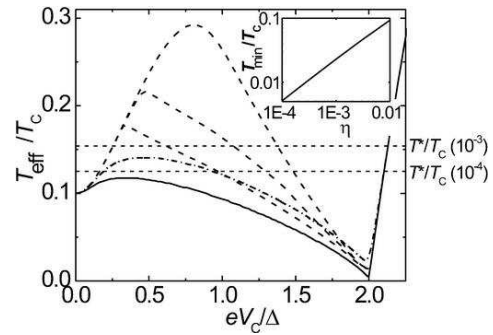}
\caption{Effective electron temperature in the cooler against the injection voltage. The dashed lines correspond to $K=0$ (extreme non-equilibrium), $K=0.1$ and $K=1.0$, from top to bottom, 
and the solid line to $K=\infty$ (quasi-equilibrium), all with $T_{\rm S}/T_{\rm C}=0.1$ and $\eta=1\cdot 10^{-4}$. The dash-dotted line is the result for quasi-equilibrium but for 
$\eta=1\cdot 10^{-3}$. The horizontal dashed lines indicate the crossover temperature $T^*$ of 
re-entrant behaviour in quasi-equilibrium for $\eta =1\cdot 10^{-4}$ and $1\cdot 10^{-3}$. The inset shows the dependence of the ultimate minimum temperature of the cooler against 
$\eta$.}\label{Figure 3}
\end{center}
\end{figure}
Figure 3 shows $T_{\rm eff}$ as a function of $v_{\rm C}$ at $T_{\rm S}/T_{\rm C}=0.1$ and $\eta = 1\cdot 10^{-4}$ for different rates $K$. In the 
collisionless limit, the rise of $T_{\rm eff}$ is largest (almost threefold)
and the maximum is reached at $v_{\rm C} \simeq 0.8$. On increasing the collision rate, the maximum $T_{\rm eff}$ gets lower and it is reached at a lower value of $v_{\rm C}$, and ultimately 
in quasi-equilibrium ($K=\infty$), the maximum is reached at $v_{\rm C} \simeq 0.4$ and its value is about 0.12$T_{\rm C}$.
The minimum temperature in quasi-equilibrium at $v_{\rm C}\simeq 2$ is given approximately by $T_{\rm min}/T_{\rm C} \simeq 2.5\eta^{2/3}$ (inset of Fig. 3). Although the experimental 
working temperatures well below 100 mK imply that $\eta < 0.01$, we cannot make a direct comparison to this theoretical result in the present device. For this we would need a thermometer 
calibration down to the lowest temperatures, the electron system should be forced to quasi-equilibrium (magnetic impurities would help), and the S reservoirs should be well thermalised. 

\begin{figure}
\begin{center}
\includegraphics[width=0.5
\textwidth]{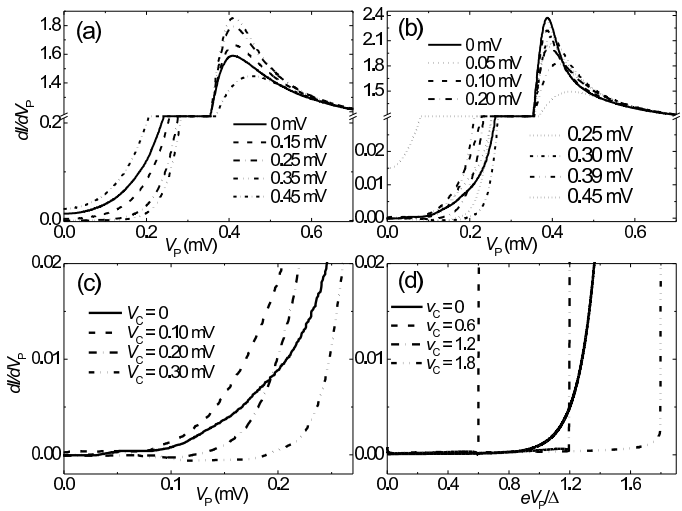}
\caption{Measured differential conductance of S1 at $T_{\rm S} = 340$ mK in (a), and at $T_{\rm S} \simeq 100$ mK in (b). Zoom-up of a few data-sets of (b) are shown in (c), and the 
corresponding theoretical lines with $K=0.05$ in (d).}\label{Figure 4}
\end{center}
\end{figure}
To assess the degree of non-equilibrium, we measured $dI/dV_{\rm P}$ at various values of $V_{\rm C}$, which allows for a semi-quantitative comparison between 
the experiment and theory (see Eq. (\ref{eq5})). In the data, especially below about 200 mK, we concentrate on the values of both $V_{\rm C}$ and $V_{\rm P}$ below the gap ($< 2\Delta/e$) 
because of the excessive heating of the reservoirs when current increases abruptly at the gap edge. The data at $T_{\rm S}=$ 340 mK in Fig. 4 (a) exhibit cooling behaviour in 
quasi-equilibrium: the conductance curves measured get first sharper monotonically on increasing $V_{\rm C}$ from 0 to 0.35 mV, whereas data at $V_{\rm C} = 0.45$ mV are more smeared, i.e. 
hotter than any other curve. All data sets conform in shape without crossing, demonstrating near-to-quasi-equilibrium behaviour. 
The data in Fig. 4 (b), taken at the base temperature of the cryostat of about 50 mK (best fit to $dI/dV_{\rm P}$ would yield $T_{\rm 
S}\simeq 100$ mK), show, on the contrary, that the energy distribution in this case deviates from the thermal one when applying bias $V_{\rm C}$. Data at $dI/dV_{\rm P} \le 0.03$ first 
indicate that the low-bias conductance becomes larger when increasing $V_{\rm C}$ from 0 up to 0.1 mV (unlike at 340 mK), where-after the curves start to get sharper, but they heavily cross 
each other in this regime. At 0.45 mV the data present significant heating again. The mere fact that the data-sets corresponding to different values of $V_{\rm C}$ criss-cross in the regime 
below 0.4 mV is a demonstration of non-equilibrium. A few data sets from 
(b) are magnified in (c), and the corresponding theoretical results, assuming $K=0.05$, have been shown in (d). The resemblance is obvious, although 
the theoretical lines show abrupt rise from $dI/dV_{\rm P}=0$ at $V_{\rm P}=V_{\rm C}$ due to the influence of the gap edge. This feature is smeared in experiment most likely because of 
noise and finite excitation level (25 $\mu$V p-p) in the measurement.

In summary, we have shown that slow electron-electron relaxation restricts the use of the concept of temperature in electron coolers at low temperatures, and that the non-zero DOS within the 
gap of the superconducting reservoirs gives rise to anomalous heating and determines the ultimate minimum temperature that can be achieved. 

We thank Academy of Finland (J.P.P.) and Institut Universitaire de France (F.W.J.H.) for financial support, and H. Courtois, L. I. Glazman, J. M. Kivioja, L. S. Kuzmin, J. E. Mooij, M. A. Paalanen, and J. N. Ullom for 
discussions.


\begin{thebibliography}{99}

\bibitem{nahum94} M. Nahum, T. M. Eiles, and J. M. Martinis, Appl. Phys. Lett. {\bf 65}, 3123 (1994).

\bibitem{leivo96} M. M. Leivo, J. P. Pekola, and D. V. Averin, Appl. Phys. Lett. {\bf 68}, 1996 (1996).

\bibitem{roukes85} M. L. Roukes, M. R. Freeman, R. S. Germain, R. C. Richardson, and M. B. Ketchen, Phys. Rev. Lett. {\bf 55}, 422 (1985).

\bibitem{pothier97} H. Pothier, S. Gu\'eron, N. O. Birge, D. Esteve, and M. H. Devoret, Z. Phys. B {\bf 104}, 178 (1997).

\bibitem{baselmans99} J. J. A. Baselmans, A. F. Morpurgo, B. J. van Wees, and T. M. Klapwijk, Nature (London) {\bf 397}, 43 (1999).

\bibitem{fisher99} P. A. Fisher, J. N. Ullom, and M. Nahum, Appl. Phys. Lett. {\bf 74}, 2705 (1999).

\bibitem{rowell76} J. M. Rowell and D. C. Tsui, Phys. Rev. B {\bf 14}, 2456 (1976).

\bibitem{andreevnote} We assume resistive enough tunnel barriers, such that Andreev reflection can be neglected. We expect significant modifications of the results presented here in the case 
of highly transparent interfaces.

\bibitem{dynes84} R. C. Dynes, J. P. Garno, G. B. Hertel, and T. P. Orlando, Phys. Rev. Lett. {\bf 53}, 2437 (1984).

\bibitem{eta} The precise value of $\eta$ is not important, as soon as it provides the dominant coupling of the island to the environment as is the case at low temperature. We use $\eta = 
1\cdot 10^{-4}$ based on the strength of sub-gap 
heating (Fig. 1 (b)), on the extrapolation of data in Ref. \cite{dynes84}, and on the ratio ($\simeq \eta$) of the conductances at low and high bias, respectively, of the measured junctions 
at low temperature.

\bibitem{vanhuffelen93} W. M. van Huffelen, T. M. Klapwijk, D. R. Heslinga, M. J. de Boer, and N. van der Post, Phys. Rev. B {\bf 47}, 5170 (1993).

\bibitem{phononestimate} As an example, we obtain $T_{\rm min} \simeq$ 2 mK with $T_{\rm S}/T_{\rm C}=0.1$ in S1, a value well below any temperature measured here.

\bibitem{collision} $\Delta$ appears in $K$ because we chose it as a suitable energy scale for writing Eq. (\ref{coll}) and to describe the whole process including injection: the true 
collision integral $I_{\rm coll}$ depends on N island properties only (not on $\Delta$).

\end{thebibliography}
\end{document}